# Explaining the apparent arbitrariness of the LDA-1/2 self-energy correction method applied to purely covalent systems

Kan-Hao Xue,[1,2] Leonardo R. C. Fonseca,[3] and Xiang-Shui Miao[1,2]

[1]School of Optical and Electronics Information, Huazhong University of Science and Technology, Wuhan 430074, China

[2]Wuhan National Laboratory for Optoelectronics, Wuhan 430074, China

[3]Center for Semiconductor Components, University of Campinas, Campinas, São Paulo 13083-870, Brazil

**Abstract**

The LDA-1/2 method expands Slater's half occupation technique to infinite solid state materials by introducing a self-energy potential centered at the anions to cancel the energy associated with electron-hole self-interaction. To avoid an infinite summation of long-ranged self-energy potentials they must be trimmed at a variationally-defined cutoff radius. The method has been successful in predicting accurate band gaps for a large number of elementary and binary semiconductors. Nevertheless, there has been some confusion regarding carbon and silicon, both in the cubic diamond structure, which require different ionizations of the valence charge, 1/2 for carbon and 1/4 for silicon respectively, to yield band gaps in agreement with experimental data. We here analyze the spatial distribution of the valence electrons of these two materials to conclude that in silicon and in carbon LDA-1/4 and LDA-1/2, respectively, must be adopted for the proper cancellation of the self-energies. Such analysis should be applied to other covalent semiconductors in order to decide which ionization to adopt for the proper correction of the self-energy.

## 1. Introduction

Despite the fundamental significance of density functional theory[1,2] (DFT) in modern computational materials science and chemistry, it is well-known that under the local density approximation[2] (LDA) or generalized gradient approximation[3,4] (GGA) the band gap is usually underestimated in semiconductors and insulators. Hybrid functionals[5,6] and the quasi-particle approach within the GW approximation[7] are usually adopted to remedy this problem at a computational cost orders of magnitude higher than standard LDA or GGA. Since the band gap is a crucial parameter in microelectronics and optoelectronics, where large supercells are frequently required for the calculation of defects, charged systems, surfaces, interfaces and grain boundaries,[8] it is of great interest to develop a method that can handle supercells with several hundred atoms while describing correctly the material's electronic structure, and in particular one that yields accurate band gaps.

There are in fact two reasons why band gaps are underestimated in LDA or GGA. First, in Mott insulators and charge transfer insulators such as some late 3d transition metal oxides, strong electron correlation splits the narrow 3d band into upper and lower Hubbard subbands, where the energy raise of the upper Hubbard subband causes an abnormally large band gap such as in NiO. The use of the single electron approximation in this case is therefore questionable since the energy of a 3d electron near the Fermi level depends on whether its orbital is already occupied by another electron or not. An efficient solution is found in the LDA+U (GGA+U) approach,[9] which possesses similar computational load as standard LDA (GGA). Nevertheless, LDA+U requires empirical parameters $U$ and $J$, and sometimes does not seem physically sound when applied to non-strongly-correlated systems. For example, in $TiO_2$ it was found that adding the $U_d$ parameter to the metal $d$ orbitals does not yield a sufficient band gap, thus an additional $U_p$ parameter must be added to the orbital O $2p$ as well.[10] However, the added $U_p$ is beyond the scope of the original LDA+U. The second reason (and most important for non-strongly-correlated materials) why the band gap is underestimated in LDA stems from its systematic limitation in dealing with excitations.[11] As the Kohn-Sham energy eigenvalue does not represent the real electron energy level, the difference between the lowest unoccupied molecular orbital (LUMO) and the highest occupied molecular orbital (HOMO) read out directly from the Kohn-Sham eigenvalues cannot be interpreted as the true band gap of the corresponding material. The correct way of relating the Kohn-Sham eigenvalue to the electronic structure is through the Slater-Janak theorem:[12]

$$\frac{\partial E}{\partial f_\alpha} = e_\alpha(f_\alpha)$$

where $E$ is the total energy of the system, $f_\alpha$ is the occupation of state $\alpha$, and $e_\alpha$ is the Kohn-Sham eigenvalue of state $\alpha$.

In 2008, Ferreira and coworkers introduced the Slater half occupation technique to the calculation of solids with DFT.[13] The so-called LDA-1/2 method can predict accurate band gaps for many elementary and binary semiconductors[14] with a computational load comparable to standard LDA. Following the half-ionization rule for accurate calculation of ionization energies of isolated atoms, Ferreira *et al.* proposed an extension to infinite solid state systems. This is achieved by building up self-energy potentials for all the anions in the solid, which are then added to their pseudopotentials in the following electronic structure calculations. Instead of calculating the self-energy potentials explicitly, they derive from the difference from the neutral atomic potentials and the corresponding half-ionized ionic potentials.[13] Because the atomic self-energy potentials are long-range they must be trimmed to be localized around the corresponding atoms only. Ferreira *et al.* introduced a polynomial-form step-function $\Theta$ such that the screened self-energy potential $V_s$ becomes:

$$V_S = V_S^0 \Theta(r) = \begin{cases} V_S^0 \left[1 - \left(\frac{r}{r_{cut}}\right)^n\right]^3, & r \leq r_{cut} \\ 0, & r > r_{cut} \end{cases}$$

where $V_S^0$ is the unscreened self-energy potential, the cutoff radius $r_{cut}$ is obtained variationally so as to maximize the band gap, and $n$ is a power index which should in principle be as large as possible in order to achieve a sharp cut.[13] There is some freedom in choosing $n$, but Ferreira *et al*. have suggested $n=8$ as a default value.

Notwithstanding its early success, until now applications of LDA-1/2 are still rare. One possible reason lies in the apparent arbitrariness in choosing LDA-1/2 or LDA-1/4. Indeed, Ferreira *et al*. argued[13] that when the excited hole generated in the valence band covers $N$ atoms, then the amount of electron charge to be stripped from each relevant atom should be 1/2$N$. Hence, in diamond silicon with covalent bonding and two atoms equally sharing each bond, it is LDA-1/4 rather than LDA-1/2 that should be employed. Nevertheless, in diamond carbon LDA-1/4 severely underestimates the band gap. Consequently, Ferreira *et al*. suggested that one should add the trimmed carbon 2$s$-1/4 and 2$p$-1/4 self-energy potentials to a single carbon atom.[13] In other words, for carbon it is still LDA-1/2 that is required to produce an accurate band gap. It is puzzling why silicon requires LDA-1/4 while carbon requires LDA-1/2 as their structures are isomorphic. It is the aim of this work to explain why these choices are indeed correct. By doing so we lay out a general framework for choosing which flavor of the self-energy correction scheme, if LDA-1/2 or LDA-1/4, to apply for other purely covalent materials.

## 2. Computational methods

DFT calculations were carried out employing the plane-wave based Vienna *Ab initio* Simulation Package[15,16] (VASP) program. The electrons considered as valence were 2$s$, 2$p$ for C, and 3$s$, 3$p$ for Si. Core electrons were approximated by projector augmented-wave pseudopotentials.[17,18] LDA was used for the exchange-correlation energy, within the parameterization of Perdew and Zunger[19] based on the quantum Monte-Carlo results by Ceperley and Alder.[20] The lattice constants of diamond carbon and silicon were fixed to their experimental values, 3.567 Å and 5.431 Å, respectively. The plane wave cutoff energy was 600 eV and a 19×19×19 Γ-centered Monkhorst-Pack $k$-mesh[21] was utilized for sampling the Brillouin zone.

LDA-1/2 and LDA-1/4 calculations were performed with 1/4 or 1/2 electron charge subtracted from the ion topmost occupied orbital (2$p$ and 3$p$ for C and Si, respectively, though sometimes also from C 2$s$ for particular comparison) and the ionic potentials generated by the ATOM program.[22]

## 3. Results and discussions

For diamond we find that effect of subtracting $e$/2 from C 2$p$ is almost equivalent to subtracting $e$/4 from C 2$s$ and $e$/4 from C 2$p$ as recommended by Ferreira *et al*.[13] In the band gap *versus* cutoff radius chart shown in Fig. 1, one can hardly discern any significant difference between the two approaches. Hence, for diamond we unambiguously adopt the LDA-1/2 scheme to C 2*p*. At the optimal cutoff radius of 2.00 bohr, the zero temperature band gap is 5.78 eV, in reasonable agreement with the experimental value 5.47 eV,[23] measured at a finite temperature (295 K). In contrast, the LDA-1/4 method predicts a considerably worse band gap of 4.94 eV employing the same optimal 2.00 bohr cutoff radius.

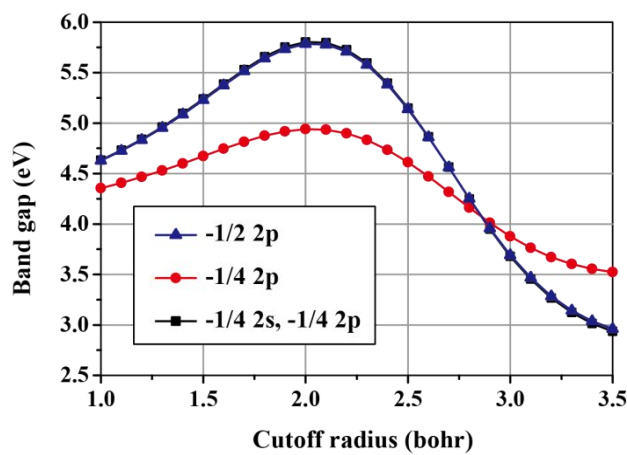

Figure 1. Calculated band gap values for diamond carbon with respect to the cutoff radius. Three ionization approaches are compared: e/2 from C 2*p* (blue triangles), e/4 from C 2*p* (red balls), and e/4 from C 2*s* together with e/4 from C 2*p* (black squares). In all cases the power index *n* adopted in the trimming function was 50 for best accuracy.

In the case of silicon, the same comparison between LDA-1/2 and LDA-1/4 now leads to a different conclusion, as illustrated in Fig. 2 where the impact of the power index *n* on the optimal cutoff radius and on the band gap is also illustrated. A larger power index leads to a sharper cut of the self-energy potential. The LDA-1/2 and LDA-1/4 band gaps for silicon are 1.95 eV and 1.19 eV, respectively, with similar optimized cutoff radii close to 3.1 bohr. LDA-1/2 fails while LDA-1/4 agrees very well with data (experimental value is well known to be 1.17 eV as extrapolated to zero temperature) as expected from Ferreira's argument for silicon.

Our results thus confirm that diamond carbon and silicon should be subject to LDA-1/2 and LDA-1/4 approaches, respectively, for accurate band gap results. To better understand this difference, in Fig. 3 we plot the ground state valence electron charge distribution (obtained from standard LDA) in silicon and carbon. We examine the valence electron distribution because the hole self-energy is most relevant at the locations with highest valence electron density. As Fig. 3 shows, there is a considerable difference where the valence electrons mostly reside: for silicon the valence electrons are located along the inter-atomic bonds, while for carbon the valence electrons are strongly concentrated around individual carbon atoms. Hence, it is natural to infer that diamond carbon does require half ionization per atom (*i.e.*, LDA-1/2) to correct the self-energy since its valence electron distribution is similar to individual carbon atoms. Figure 1 shows that the carbon band gap starts to decrease after the cutoff radius reaches 2.00 bohr, a relatively small value like those found in many ionic compounds,[13] when the anion self-energy potentials from nearest neighbors start to touch each other.

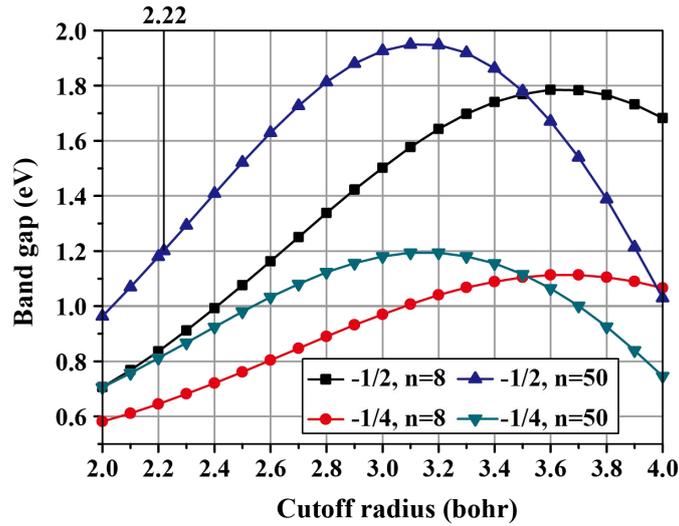

Figure 2. Calculated band gap values for diamond silicon with respect to the cutoff radius. Two ionization approaches from Si 3p and two trimming factors are compared: e/2 and *n*=8 (black squares), e/2 and *n*=50 (blue triangles), e/4 and *n*=8 (red balls), as well as e/4 and *n*=50 (light blue inverted triangles).

On the other hand, in silicon the valence electrons are shared by the adjacent silicon atoms. Consequently, employing either LDA-1/2 or LDA-1/4 one finds that the band gap continues to increase until around 3.1 bohr (Fig. 2), which is approximately 3/4 of the Si-Si bond distance (4.44 bohr). Going beyond the 3/4 bond length, which means going beyond the region where the valence electrons mostly reside, the band gap starts to decrease. Therefore in LDA-1/2 one inevitably takes into account the hole self-energy twice for the valence electrons which are symmetrically separated by the bond center.

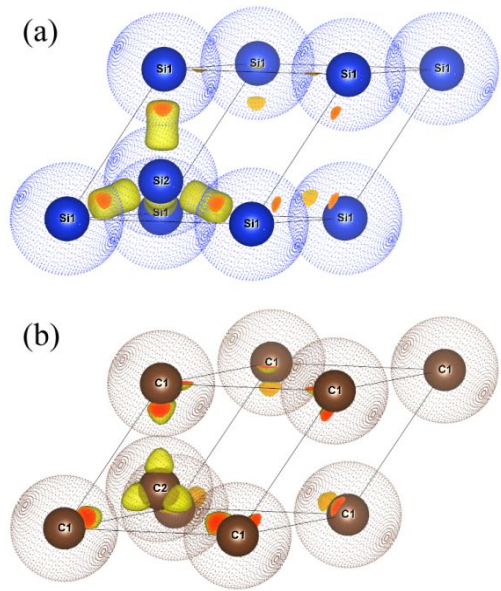

Figure 3. Valence electron charge contour shown by the yellow surfaces for (a) silicon and (b) carbon.

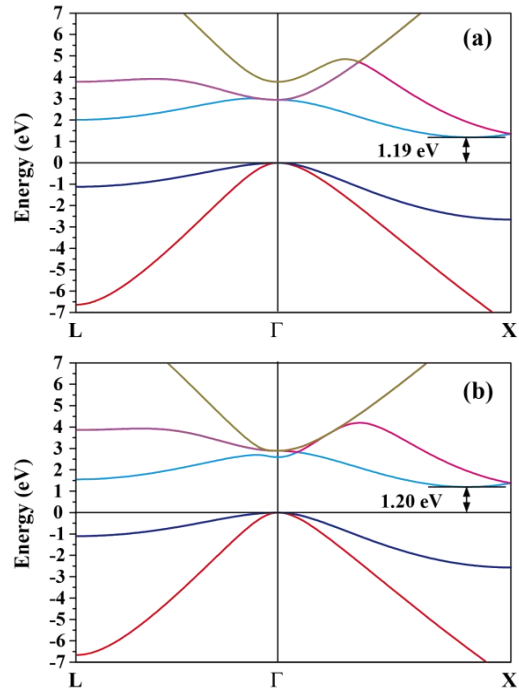

Figure 4. Electronic band structures for silicon calculated by (a) LDA-1/4 using a 3.10 bohr cutoff radius obtained from the standard variational procedure, and (b) LDA-1/2 at half bond length, i.e., manually selecting a 2.22 bohr cutoff radius. The top of the valence bands are set to zero energy in both cases.

If the problem with LDA-1/2 in Si is caused by overlapping self-energy potentials of neighboring Si atoms, then one could consider restricting the cutoff radius for silicon, to avoid that it exceeds half bond length, i.e., 2.22 bohr. For $n$=50 such cutoff-radius-restricted LDA-1/2 yields a band gap of 1.20 eV, very close to the LDA-1/4 result, 1.19 eV. However, as Fig. 4 shows, LDA-1/4 yields the correct band structure for silicon but LDA-1/2 at half bond length predicts a wrong band ordering at the Γ-point. Indeed, in LDA-1/4 the first and second conduction bands are degenerated at Γ but the third conduction band lies above, consistent with the previous well-acknowledged results.[24] In LDA-1/2 at half bond length, the second and third conduction bands are degenerated at Γ but the first conduction band lies slightly below. Hence, the unrestricted optimization of the cutoff radius is necessary in Si, thus requiring a reduction of the ionization charge from e/2 to e/4 as originally proposed by Ferreira *et al*.[13]

In the case of carbon the optimal cutoff radii for LDA-1/4 and LDA-1/2 are the same, *i.e.*, 2.00 bohr. Since the C-C bond length is 2.92 bohr, this cutoff radius is not sufficiently large as to include the valence electrons surrounding the neighboring carbon atoms. Hence, LDA-1/2 should be employed in this case.

## 4. Conclusion

The use of the LDA-1/2 technique for the calculation of accurate electronic structures in simple covalent semiconductors has previously faced the apparent contradicting choice of LDA-1/2 for diamond carbon and LDA-1/4 for diamond silicon. In this paper the issue has been clarified by a careful examination of the spatial distribution of valence electrons. In silicon the valence electrons accumulate near the bond center, thus LDA-1/4 rather than LDA-1/2 is necessary since the self-energy correction applied to every Si atom in the lattice is double counted due to the overlap of the self-energy potentials. In carbon the valence electrons are hold tightly to the carbon atoms, resembling the isolated carbon atoms. In this case, LDA-1/2 is necessary as there is no overlap of self-energy potentials from neighboring atoms. Therefore, one must consider the cutoff radius of the self-energy potential to determine whether to employ LDA-1/4 or LDA-1/2 in covalent semiconductors.

## Acknowledgement

This work was funded by the National Natural Science Foundation of China under grant No. 61376130.